\documentclass[twocolumn,prd,superscriptaddress,nofootinbib]{revtex4-1}
\usepackage{amsfonts,amsmath,amssymb,mathrsfs}
\usepackage{hyperref}
\usepackage{color}
\usepackage{soul}
\usepackage{amsmath}
\usepackage{graphicx}  
\usepackage{dcolumn}   
\usepackage{bm}        
\usepackage[table]{xcolor}
\usepackage[english]{babel}
\usepackage{comment}
\def\a{\alpha}

\def\r{\rho}
\def\s{\sigma}
\def\t{\tau}
\def\m{\mu}
\def\n{\nu}
\def\k{\kappa}
\def\th{\theta}
\def\g{\gamma}\def\G{\Gamma}
\def\L{t}\def\l{V}
\def\D{\Delta}
\def\la{\langle}
\def\ra{\rangle}
\def\o{\omega}\def\O{\Omega}
\def\d{\delta}
\def\p{\partial}

\def\oxthree{{\cal O}(x^3) }

\def\half{\textstyle{\frac{1}{2}}}

\def\bdoc{\begin{document}}
\def\edoc{\end{document}}
\def\bea{\begin{equation}}
\def\eea{\end{equation}}

\def\beq{\begin{eqnarray}}
\def\eeq{\end{eqnarray}}
\def\be{\begin{eqnarray}}
\def\ee{\end{eqnarray}}
\def\ben{\begin{enumerate}}
\def\een{\end{enumerate}}
\def\la{\langle}
\def\ra{\rangle}
\def\a{\alpha}
\def\g{\gamma}\def\G{\Gamma}
\def\d{\delta}\def\D{\Delta}
\def\e{\epsilon}
\def\z{\zeta}

\def\th{\theta}
\def\k{\kappa}
\def\l{t}
\def\m{\mu}
\def\n{\nu}
\def\o{\omega}
\def\p{\pi}
\def\r{\rho}
\def\s{\sigma}
\def\t{\tau}
\def\L{{\cal L}}
\def\S{\Sigma }
\def\gsim{\; \raisebox{-.8ex}{$\stackrel{\textstyle >}{\sim}$}\;}
\def\lsim{\; \raisebox{-.8ex}{$\stackrel{\textstyle <}{\sim}$}\;}
\def\gtrsim{\gsim}
\def\lessim{\lsim}
\def\loc{{\rm local}}
\def\vm{v_{\rm max}}
\def\bh{\bar{h}}
\def\del{\partial}
\def\nab{\nabla}
\def\half{{\textstyle{\frac{1}{2}}}}
\def\fourth{{\textstyle{\frac{1}{4}}}}

\def\bD{{\bf D}}
\def\bE{{\bf E}}
\def\bF{{\bf F}}
\def\bB{{\bf B}}
\def\bP{{\bf P}}
\def\bV{{\bf v}}
\def\bv{{\bf v}}
\def\bx{{\bf x}}
\def\by{{\bf y}}
\def\bz{{\bf z}}
\def\ba{{\bf a}}
\def\bd{{\bf d}}
\def\bs{{\bf s}}
\def\bn{{\bf n}}
\def\bp{{\bf p}}

\def\O{\Omega}

\def\br{{\bf r}}
\def\bnab{{\bf \nab}}

\def\tE{\tilde{E}}
\def\tL{\tilde{L}}
\def\Horava{Ho\v{r}ava }

\def\oxtwo{\mathscr{O}\left(x^2\right)}
\def\oxthree{\mathscr{O}\left(x^3\right)}
\def\oxfour{\mathscr{O}\left(x^4\right)}
\def\oxfive{\mathscr{O}\left(x^5\right)}
\def\LL{\text{Lanczos-Lovelock}}

\def\ph{\phantom}

\begin{document}
\title{Constraining Topological Gauss-Bonnet Coupling from GW150914}
\author{Kabir Chakravarti}
\email{kabir.c@iitgn.ac.in}
\affiliation{Indian Institute of Technology, Gandhinagar, Gujarat 382355, India.}
\affiliation{CEICO, FZU-Institute of Physics of the Czech Academy of Sciences, Na Slovance 1999/2, 182 21 Prague 8, Czech Republic}
\author{Rajes Ghosh}
\email{rajes.ghosh@iitgn.ac.in }
\affiliation{Indian Institute of Technology, Gandhinagar, Gujarat 382355, India.}
\author{Sudipta Sarkar}
\email{sudiptas@iitgn.ac.in}
\affiliation{Indian Institute of Technology, Gandhinagar, Gujarat 382355, India.}

\begin{abstract}
\noindent
Recent gravitational wave observation based on the data from GW150914 has confirmed Hawking's area theorem and estimated the increase in total horizon area during a merger process of two Kerr black holes. We use this result and the validity of the second law to obtain the first observational bound on the 4D topological Gauss-Bonnet coupling as $\gamma \lesssim 2.804^{+7.946}_{-1.169} \times 10^{9}\,  m^2$ with $95 \%$ credibility. 
\end{abstract}
 
\maketitle 
\textit{Introduction.}\textemdash Black holes (BHs) are arguably the simplest among all astrophysical objects found in nature. Yet, they often provide profound insights into the nature of gravity and the structure of spacetime. The laws governing their mechanics exhibit intriguing resemblance with that of ordinary thermodynamics \cite{Bardeen,Bekenstein:1972,Bekenstein:1973}. In particular, Hawking's area theorem ensures, akin to the second law of thermodynamics, that area of the event horizon cannot decrease in any classical process \cite{Hawking:1971vc}. Also, the study of the quantum fields in a BH spacetime allows us to associate a notion of temperature with its event horizon \cite{Hawking}. This Hawking temperature of a stationary BH is proportional to the surface gravity that remains constant throughout the event horizon leading to the zeroth law of BH thermodynamics. Moreover, the temperature-surface gravity proportionality constant fixes the area-entropy relationship via the first law. Thus, for a BH of mass $M$ and horizon area $A$, the Hawking temperature $T_H$ and entropy $S$ are given by,

\beq
T_H = \frac{1}{8 \pi M}\, ,\,\, \textrm{and} \,\,\, S = \frac{A}{4},
\eeq

\noindent
in natural units $c=G=k_B=\hbar=1$. Although these laws were originally developed for general relativity (GR) alone, there has been a surge of research towards extending their applicability beyond GR \cite{Visser:1993nu, Jacobson:1993xs, Wald:1993nt, Iyer:1994ys, Jacobson:1994qe, Wald:1999vt, Chatterjee:2011wj, Sarkar:2013swa, Bhattacharjee:2015qaa, Wall:2015raa, Wall:2018ydq, Sarkar:2019xfd, Ghosh:2020dkk}.\\

\noindent
In GR, the proof of the second law requires the area theorem which is crucially based on the validity of Einstein's equations \cite{Hawking:1971vc, Bardeen}. To calculate the rate of change of horizon area, one uses the null version of the Raychaudhuri's equation that, along with the field equations, relates the expansion of the null generators of the event horizon with the matter content $T_{ab}$. Then, by assuming the null energy condition on matter, it is straightforward to show that the expansion always remains positive provided the horizon generators are future-complete and free from caustic formation in the future.\\

\noindent
Now, consider the process in which two BHs merge together to form another BH. This event will generate bursts of gravitational radiation carrying away energy out of the system. The mass of the final BH is determined by the difference of the total initial mass and the amount of gravitational radiation. A priori, it seems energetically possible to have arbitrarily large amounts of radiation just by tuning the mass of the final BH to arbitrarily small values. However, the area theorem restricts the maximum amount of gravitational radiation that can be extracted from the merger by imposing an upper bound on the efficiency of this process \cite{Hawking:1971tu}. Therefore, the verification of the area increase during the merger of two BHs is an important test of GR.\\

\noindent
Recently, the authors of Ref.\cite{Isi:2020tac} have analysed the merger data of GW150914 and claimed the confirmation of the area theorem. This result involves independent analysis of the inspiral and ringdown phases of the GW signal, and measuring the change in the total horizon area $\Delta A$ \footnote{The test of area theorem was also proposed previously in Ref.\cite{Cabero:2017avf}, by separately fitting the early inspiral and final ringdown stages, thereby avoiding the merger phase where the assumption of a stationary Kerr BH is not valid.}. However, the analysis in Ref.\cite{Kastha:2021chr} of the time domain binary BH evolution points out possible errors that may lead to overstating the confidence with which the area theorem is confirmed in Ref.\cite{Isi:2020tac}. Also, a recent work \cite{Cotesta:2022pci} questions the validity of the overtone detection used in Ref.\cite{Isi:2020tac}. Nevertheless, all these studies suggest that the area of the final BH is always larger than the total area of the two initial coalescing BHs, in accordance with the area theorem. \\

\noindent
If the BH entropy is proportional to the horizon area, these results verify the thermodynamic nature of BH horizons as well.  Note that these observations only verify the global version of the area theorem, i.e., the area of the final BH is larger than the total area of the two initial ones. The area theorem actually asserts a stronger result that the instantaneous change of area is always positive too. Moreover, the present GW data is consistent with the Kerr paradigm, namely the astrophysical BHs generating gravitational waves through merger are described by the vacuum Kerr solutions of GR \cite{Isi:2019aib}.\\

\noindent
However, there is still a possibility that the theory of GR may receive corrections due to some unknown UV-physics. This is quite plausible given the fact that the quantum GR is perturbatively non-renormlizable and may only make sense as an effective theory. Then, we expect various higher curvature terms to arise in the gravitational action at relevant length scales. In four dimensions (4D), the most general higher curvature action up to quadratic order is given by,

\beq \label{ac}
{\cal A} = \frac{1}{16 \pi \, }\int \sqrt{-g}\,  \, d^4 x \left( R +  \alpha\, R^2 + \beta\, R_{ab} R^{ab} \right.\\ \nonumber  \left.+\, \gamma\, R_{abcd} R^{abcd} \right)\ .
\eeq

\noindent
This action contains new dimensionful parameters $\alpha, \beta, \text{and}\, \gamma$ denoting the length scales at which the corresponding higher curvature terms are important. The dynamics of such modified theories may substantially differ from GR. In particular, the area-entropy proportionality may get violated due to the presence of such higher curvature terms in the action \cite{Visser:1993nu, Jacobson:1993xs, Wald:1993nt, Iyer:1994ys}. \\

\noindent
Next, one can rewrite this action as follows

\beq
{\cal A} = \frac{1}{16 \pi \, }\int \sqrt{-g}\,  d^4 x \left( R +  \alpha\, R^2 + \beta\, R_{ab} R^{ab} \right.\\ \nonumber  \left. +\, \gamma\, {\cal L_{GB}} \right) \label{action}\ ,
\eeq

\noindent
where we have redefined the dimensionful parameters and used the expression for the Gauss-Bonet (GB) term, ${\cal L_{GB}} = R^2 - 4 R_{ab} R^{ab} + R_{abcd} R^{abcd}$. In 4D, this term is topological in nature and thereby, can not influence the dynamics. Also, the Kerr geometry remains a solution of this theory \cite{Psaltis:2007cw}, though it may not be the unique axisymmetric asymptotically flat vacuum solution unlike GR. \\

\noindent
There are many known observational and experimental constraints on the higher curvature couplings $\alpha$ and $\beta$. For instance, the study of possible violation of the inverse square law imposes strict constraints on these terms \cite{Kapner:2006si,Berry:2011pb}. Moreover, the study of the delay between gravitational and electromagnetic signal from coalescing neutron stars can lead to a bound on the parameter $\beta$ \cite{Ghosh:2019twk}. However, the coefficient $\gamma$ cannot be constrained by any such observations/experiments as the corresponding term does not change the field equation. As a result, currently, there is no bound on this coefficient and any value of $\gamma$ is allowed.  \\

\noindent
Although the classical field equations are oblivious to the GB topological term, nevertheless there are some important consequences of this term. For example, the inclusion of this term with $\gamma > 0$ in the gravitational action can increase the instability of 4D de Sitter spacetime by allowing nucleation of BHs \cite{Parikh:2009js}. However, the most striking effect of this topological term is on the second law of BH mechanics. In fact, this term may lead to the violation of the BH entropy increase during a merger process.\\

\noindent
In this letter, we show that the GW150914 observation of BH area increase reported in Refs.\cite{Isi:2020tac} and \cite{Kastha:2021chr} along with the validity of the BH second law can significantly constrain the topological GB term by providing a stringent upper bound on the coefficient $\gamma$. As per our knowledge, this is the first such bound on the topological term in 4D. This bound can also be utilized to put effective constraints on the phenomena that are sensitive to $\gamma$ such as nucleation of BHs \cite{Parikh:2009js}.\\

\textit{Second Law \& Topological GB Term.}\textemdash  We can calculate the entropy of a stationary BH solution of the theory described by the action in Eq. (\ref{action}) using Wald's formula to obtain \cite{Wald:1993nt, Jacobson:1993xs, Liko},

\beq \label{S}
S = \frac{1}{2}\int \left ( \frac{1}{2} +  \alpha R - \beta R_{ab} k^a l^b +  \gamma\,\, {}^{(2)}R \right) dA\ .
\eeq

\noindent
Here, $k^a$ and $l^a$ are the null generator, and the auxiliary null vector of the horizon, respectively. They are related by the normalization $k^a l_a = -1$. Moreover, ${}^{(2)}R$ is the intrinsic curvature of the horizon cross-section, which is a compact two dimensional surface and is a topological sphere. The integration measure $dA$ is the infinitesimal area element of this cross section. \\

\textit{Entropy change in a merger process.}\textemdash Let us now consider the process of the merger of two BHs. Initially, the BHs can be well approximated by two isolated Kerr solutions with masses and spins $(M_i, a_i)$, with $i = 1,2$. Also, the final BH after the merger is also well described by the vacuum Kerr geometry of parameters $(M_f, a_f)$. As a result, much before and after the merger, the Ricci tensor (therefore, the Ricci scalar) vanishes and only the last term in the RHS of Eq. (\ref{S}) survives apart from the GR contribution. However, since the horizon cross-sections are two dimensional compact surfaces, the last term is actually a topological constant. In fact, due to the GB theorem, it is proportional to the Euler number $\chi$ associated with the surface. Thus, for the purpose of studying the global second law, we may write the BH entropy as \cite{Liko},

\beq
S = \frac{A}{4} + 2 \pi \gamma \chi\ .
\eeq

\begin{figure} [h!]
\begin{center}
\includegraphics[width=7cm]{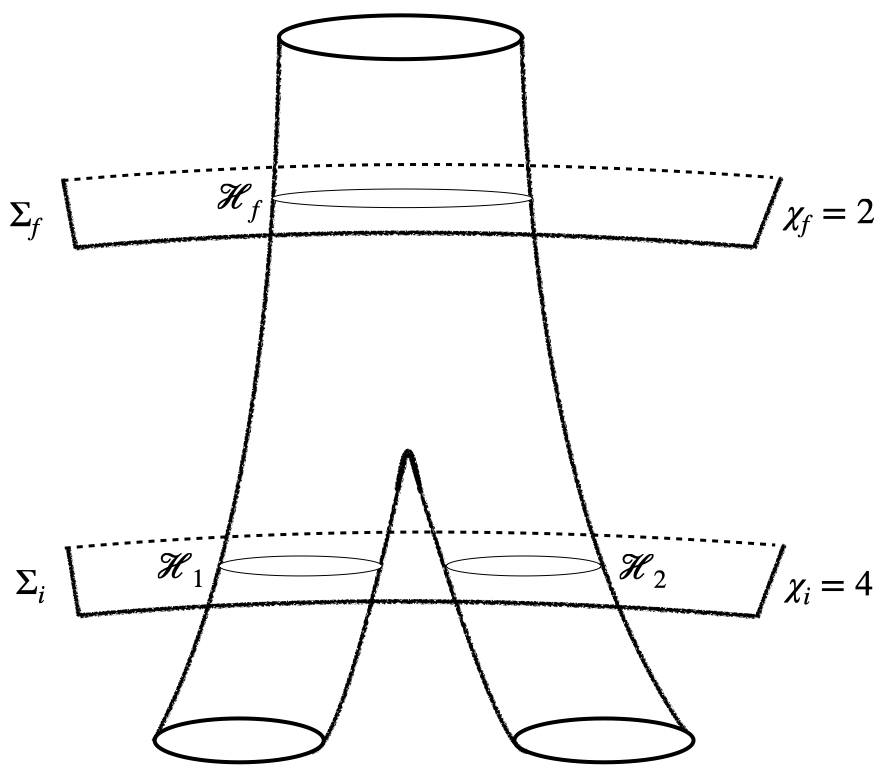}
\end{center}
\caption{Formation of the final Kerr BH with horizon slice $\mathcal{H}_f$ by merging two initial Kerr BHs with horizon slices $\mathcal{H}_1$, and $\mathcal{H}_2$. The Euler number changes discontinuously from the initial to final Cauchy slice, $\Delta \chi = 2-(2+2)=-2$. }
\label{merger}
\end{figure}

\noindent
One expects this entropy to increase during a dynamical process. If the horizon cross-sections are topological spheres, as in the case of Kerr geometry, we have $\chi = 2$. So, it seems that the last term of the above equation is a constant and would not contribute to any dynamical change of the BH spacetime. However, this is actually a hasty conclusion that is only valid as long as the topology of the horizon cross-section remains unchanged. Once we consider the process of the merger of two BHs, the initial horizon slice $\Sigma_i$ has the topology of a disjoint union of two 2-spheres, whereas the final horizon slice $\Sigma_f$ has the topology of a single 2-sphere, see Fig. [\ref{merger}]. Therefore, we have a transition between different topologies, which occurs at the point of the merger. The location of this merger point depends on the choice of foliation but if the classical second law holds analogous to the area increase theorem of GR, the entropy should be increasing no matter what foliation is used. Comparing the slices just before and after the merger, the area changes continuously, whereas the Euler number suffers an instantaneous jump at the exact moment of topology change. \\

\noindent
Even if we consider only the global version of the second law, comparing the entropy of the initial to the final slices, we see that the total change of entropy is,

\beq
\Delta S = \frac{\Delta A}{4} + 2 \pi \gamma \Delta \chi\ ,
\eeq

\noindent
with $\Delta \chi = 2 - 4 = -2$, for our case. This, in turn, implies that

\beq
\Delta S = \frac{\Delta A}{4} - 4 \pi \gamma  \label{entch}\ .
\eeq

\noindent
The above relationship shows that even if the change in horizon area $\Delta A >0$, we can choose values of the coefficient $\gamma$ which can make $\Delta S < 0$ \cite{Liko, Sarkar:2010xp}, leading to the violation of the second law. Note, this is a global violation of the second law that suggests the entropy of the final BH is less than the total initial entropy. \\

\noindent
We can also consider the local violation of the second law. In case of pure GR, the event horizon area increases at every instant. In the presence of the topological GB term, while the area increases continuously, the Euler number changes discontinuously at the merger. Thus, there is always a slice on which $\Delta S$ is negative leading to an instantaneous decrease of the entropy irrespective of the values of GB coefficient $\gamma$. Even a very small value of $\gamma$ may lead to such a decrease \cite{Sarkar:2010xp}. Nevertheless, we can always argue that Wald's formula for BH entropy may not be applicable for a violent non-stationary scenario such as BH merger. After all, the original derivation of this formula requires a stationary BH. However, this argument cannot be used if there is a global violation of the second law, as the initial and the final time slices, the metric closely resemble a stationary Kerr spacetime.\\

\noindent
Moreover, in the work Ref. \cite{Chatterjee:2013daa}, it is argued that if we consider the 4D Einstein-Gauss-Bonnet gravity as an effective theory, then the violation of the second law requires a regime where semi-classical approximation brakes down. While this argument indeed has some merit, it does not rule out the violation beyond the effective theory assumption. Note, we can use any value of the parameter $\gamma$, without modifying the classical equations of motion. Therefore, it is important to find the constraint of this parameter so that there is no such violation of the second law. In fact, in Ref. \cite{Sarkar:2010xp}, it is shown that one can use the entropy decrease for the 4D topological GB term and create situation that can lead to the decrease of entropy even in higher dimensional Lovelock gravity. This can be achieved via a Kaluza-Klein compactification of the original 5D spacetime into a 4D spacetime. \\

\textit{Bound on the GB coupling constant.}\textemdash Let us now return to Eq. (\ref{entch}) that suggests a bound on the GB coupling constant $\gamma$ so that the entropy increases and second law of BH thermodynamics remains valid,

\beq \label{ineq}
\gamma < \frac{\Delta A}{16 \pi}  \label{bound} := \gamma_{max}\ .
\eeq

\noindent
Therefore, if we have a measure of the total change of area during a merger of two Kerr BHs, we can immediately obtain an upper bound on the parameter $\gamma$. This bound is necessary to ensure the validity of the second law in the presence of the 4D topological GB term. \\

\noindent
Here we have assumed that the coefficient $\gamma$ to be positive. This is motivated from the result that for $D>4$, the coefficient of the GB term cannot be negative \cite{Cheung:2016wjt}. Then, we can immediately impose a lower bound on $\gamma$ by demanding positivity of entropy in all dimensions. In fact, for $\gamma < 0$, the second law can be violated when a black hole forms from collapse, at the instant that the horizon first appears \cite{Liko, Sarkar:2010xp}. All these arguments justify our choice $\gamma > 0$.\\

\noindent
We may use the analysis in Ref.\cite{Isi:2020tac} of the GW data from the black hole merger event GW150914 to obtain the area change and employ it to find the upper bound $\gamma_{max}$ on the GB coupling. However, since the data and the corresponding analysis of Ref.\cite{Isi:2020tac} are not yet publicly available, one might get only a rough point estimate, $\Delta A/ A_0 < 0.60$ from the $[220]$ mode. Then, using Eq.(\ref{ineq}), it directly implies: $\left(\gamma_{max}/A_0\right) < 0.012$.

\begin{figure} [h!]
\begin{center}
\includegraphics[width=9cm]{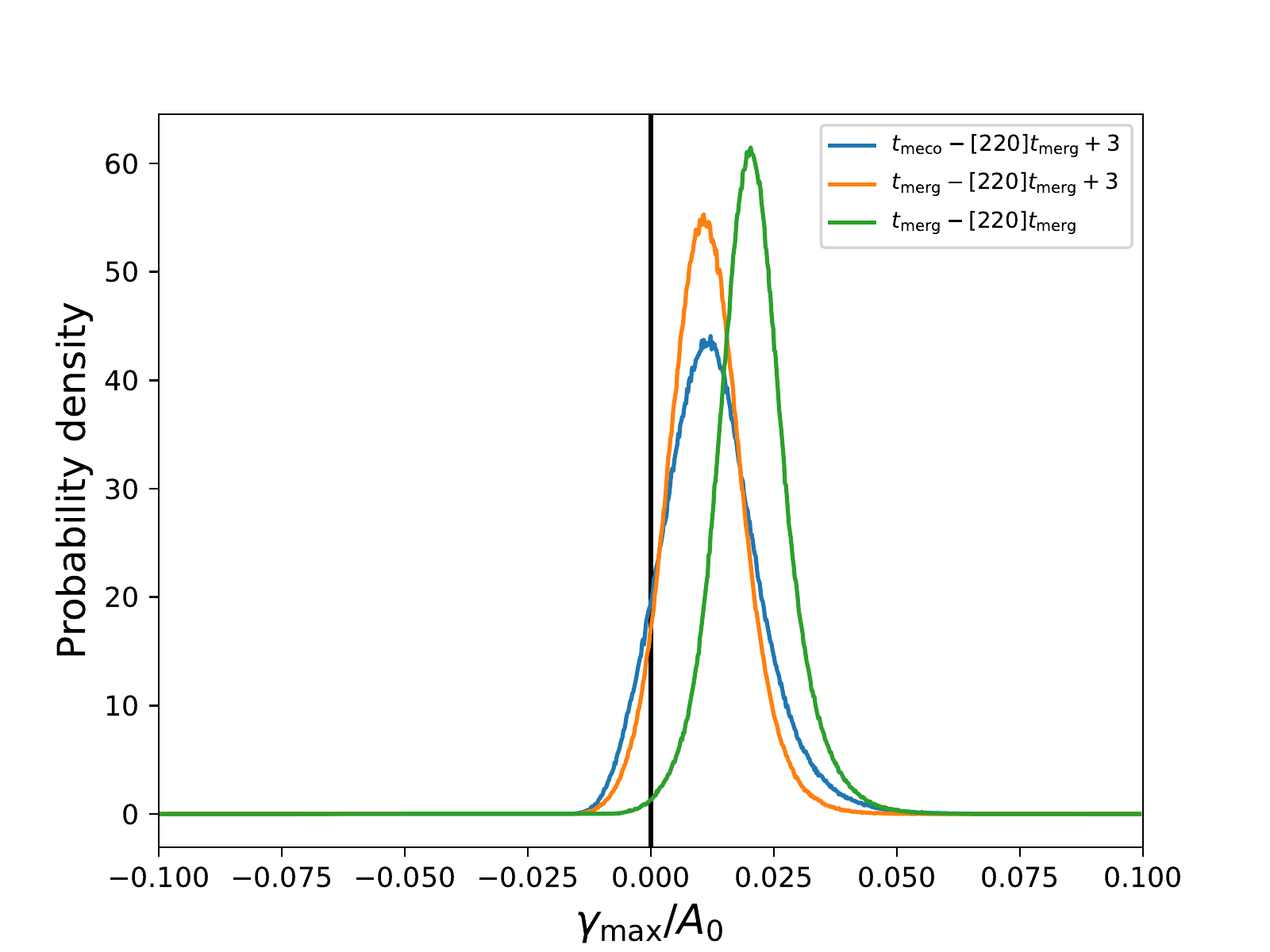}
\end{center}
\caption{Probability density of the dimensionless quantity $\left(\gamma_{max}/A_0\right)$ using data from Ref.\cite{Kastha:2021chr}. The region to the left of the black vertical line violates the area law.}
\label{gamma2}
\end{figure}

\noindent
However, to infer a much better estimation for $\gamma_{max}$ we now consider the probability distribution of $\left(\Delta A/A_0\right)$ that is provided in Ref.\cite{Kastha:2021chr}. We use their result to obtain the corresponding probability distribution for the dimensionless parameter $\left(\gamma_{max}/A_0\right)$, where $A_0$ is the total area of the two initial Kerr BHs. Our result is depicted in Fig.[\ref{gamma2}].\\

\noindent
Depending on the choice of the model, see Table \ref{table1}, we obtain the following upper bounds on $\left(\gamma_{max}/A_0\right)$ at $95 \%$ credibility $(2\sigma)$. Here, the abbreviations ``meco" and ``merg" refer to minimum energy circular orbit and merger, respectively.

\renewcommand{\arraystretch}{2.5} 
\begin{table}[h!]
\begin{tabular}{|c|c|}
\hline
~\bf{Estimates}~ & ~\bf{Models}~ \\
\hline
$0.012^{+0.034}_{-0.005}$ & $(\text{NRSur7dq4 before}\, t_{\mathrm{meco}})$ \\ & Damped sinusoid $[220](\text{after}\,  t_{\mathrm{merg}}+3$ ms)\\ 
\hline
$0.011^{+0.027}_{-0.003}$ & $(\text{NRSur7dq4 before}\, t_{\mathrm{merg}})$ \\ & Damped sinusoid $[220](\text{after}\,  t_{\mathrm{merg}}+3$ ms)\\
\hline
$0.020^{+0.006}_{+0.037}$ & $(\text{NRSur7dq4 before}\,  t_{\mathrm{merg}})$ \\ & Damped sinusoid $[220](\text{after}\,  t_{\mathrm{merg}})$ \\
\hline
\end{tabular}
\caption{Upper bounds on $\left(\gamma_{max}/A_0\right)$ with $95\%$ credibility region from different inspiral-merger-ringdown models as suggested in Ref.\cite{Kastha:2021chr}.}
\label{table1}
\end{table}

\noindent
We observe that while the first two Models are in good agreement with each other and also with the point estimation from Ref.\cite{Isi:2020tac}, the last Model peaks at a value of $\left(\gamma_{max}/A_0\right) \approx 0.020$ that has about $80 \%$ deviation from the other ones. This anomaly is due to inaccuracies arising from the assumption that the damped sinusoidal quasi-normal modes initiate right after the merger at $t=t_{\textrm{merg}}.$ For this reason, we restrict ourselves to use only the first two Models. Hence, we obtain an approximate upper bound: $\gamma \lesssim \gamma_{max} \approx 2.804^{+7.946}_{-1.169} \times 10^{9}\, m^2$ at $95 \%$ credibility. We also note that the peak value of the above estimate agrees quite well with the point estimate from Ref.\cite{Isi:2020tac}.\\

\noindent
To obtain this bound, we have neglected the spins of the initial black holes and treated them as Schwarzschild black hole of masses $35.8 M_\odot$ and $29.1 M_\odot$. This is because the estimated values of the spins of the initial black holes \cite{LIGOScientific:2016vlm} do not change the order of magnitude of the bound. \\

\noindent
Let us place our estimate for $\gamma_{max}$ in the perspective of the known bounds on similar higher curvature couplings such as $\alpha R^2$ and $\beta R_{ab} R^{ab}$. The Newtonian limit of these higher curvature modifications introduce Yukawa-like departures in the gravitational potential that can be verified by E\"{o}t-Wash experiment. Such an experiment puts an upper bound on $\alpha$ as $2 \times 10^{-9} \, m^2$ \cite{Kapner:2006si, Berry:2011pb}. This is possibly the best bound on the non-topological higher curvature theory from a local table-top experiment. Regarding astrophysical observations, Gravity Probe B experiment also imposes a bound, $\alpha \lesssim 5 \times 10^{11} \, m^2$ \cite{Naf:2010zy}. Similarly, planetary precession rates put the upper bound of $2.4 \times 10^{18} \, m^2$ on the coupling of the $R^2$ gravity \cite{Berry:2011pb}. The study of Hulse-Taylor binary pulsar data leads to a bound of $ 1.1 \times 10^{16} \, m^2$ \cite{Vilhena:2021bsx}. There is a weak constraint on the coefficient of the $R_{ab} R^{ab}$ from time delay between GW170817 and GRB 170817A as $\beta \lesssim 10^{36}\, m^2$ \cite{Ghosh:2019twk}. Similar bounds on both $\alpha$, and $\beta$ are known from the study of GW generated from binary inspirals \cite{Kim:2019sqk}. \\

\noindent
Note that these bounds on the higher curvature terms like $R^2$ and $R_{ab} R^{ab}$ are possible to obtain from analysis of the classical gravitational dynamics. The topological GB term, in contrast, cannot be constrained by any such means. Therefore, to the best of our knowledge, this is the only work that provides an observational constraint on such a term. Moreover, in terms of magnitude, our bound is considerably better than the previous astrophysical bounds on other higher curvature couplings.\\

 \noindent
In conclusion, the observation of the area increase during a BH merger allows us to bound the hitherto unconstrained parameter $\gamma$, namely the coefficient of the topological GB term in 4D. In future, if we detect a BH merger process having lower efficiency of generating GWs, we will be able to improve this bound significantly. Also, note that the values of $\gamma$ leading to violations of the global second law remain unconstrained by our analysis.\\

\noindent
\textit{Acknowledgement.}\textemdash We sincerely thank the authors of Ref.\cite{Kastha:2021chr} for providing us with the data of the probability distribution of the quantity $\left(\Delta A/A_0\right)$. We also acknowledge the anonymous referee for helpful suggestions that improve the content and presentation of the paper. K. C acknowledges support from
the Czech Academy of Sciences under Project No. LQ100102101. The research of R.G. is supported by the Prime Minister Research Fellowship (PMRF-192002-120), Government of India. The research of S.S. is supported by the Department of Science and Technology, Government of India under the SERB CRG Grant (CRG/2020/004562). We thank Anand Sengupta and Soumen Roy for helpful discussions and comments.

\end{document}